\documentclass{ws-ijmpcs}

\newcommand{\bmath}{\begin{displaymath}}
\newcommand{\emath}{\end{displaymath}}
\newcommand{\bite}{\begin{itemize}}
\newcommand{\eite}{\end{itemize}}

\renewcommand{\P}{\mathcal{P}}
\newcommand{\eps}{\varepsilon}

\newcommand{\bell}{\mathbf{\ell}}

\newcommand{\D}{\mathfrak{D}}

\newcommand{\V}{\mathcal{V}}

\newcommand{\E}{{\mathcal{E}}}

\newcommand{\bx}{\mathbf{x}}

\renewcommand{\O}{\mathcal{O}}

\newcommand{\bel}[1]{\begin{equation}\label{#1}}
\newcommand{\bal}[1]{\begin{eqnarray}\label{#1}}
\newcommand{\ee}{\end{equation}}
\newcommand{\ea}{\end{eqnarray}}

\newcommand{\Tr}{{\rm Tr}}

\newcommand{\tE}{{\tilde\E}}
\newcommand{\tphi}{{\tilde\phi}}
\newcommand{\set}[1]{{\{#1\}}}

\newcommand{\fig}[1]{fig.~\ref{#1}}

\newcommand{\equ}[1]{eq.~(\ref{#1})}
\newcommand{\Equ}[1]{Eq.~(\ref{#1})}

\begin{document}

\markboth{Martin Schaden}
{Irreducible Scalar Many-Body Casimir Energies}

%
\catchline{}{}{}{}{}
%

\title{IRREDUCIBLE SCALAR MANY-BODY CASIMIR ENERGIES:\\THEOREMS AND NUMERICAL STUDIES}

\author{MARTIN SCHADEN}

\address{Physics Department, Rutgers University, 101 Warren Street\\
Newark, New Jersey 07102, United States of America\\
mschaden@rutgers.edu}

\maketitle

\begin{history}
\received{Day Month Year}
\revised{Day Month Year}
\end{history}

\begin{abstract}
We define irreducible $N$-body spectral functions and Casimir energies and consider a massless scalar quantum field interacting locally by positive potentials with classical objects. Irreducible $N$-body spectral functions in this case are shown to be conditional probabilities of random walks. The corresponding irreducible contributions to scalar many-body Casimir energies are finite and positive/negative for an odd/even number of objects. The force between any two finite objects separable by a plane is always attractive in this case.   Analytical and numerical world-line results for the irreducible four-body Casimir energy of a scalar with Dirichlet boundary conditions on a tic-tac-toe pattern of lines are presented. Numerical results for the irreducible three-body Casimir energy of a massless scalar satisfying Dirichlet boundary conditions on three intersecting lines forming an isosceles triangle are also reported. In both cases the symmetric configuration (square and isosceles triangle) corresponds to the minimal irreducible contribution to the Casimir energy.   
\end{abstract}

\ccode{PACS numbers:11.10.-z,11.10.Gh,11.10.Jj, 11.80.Jy, 31.15.xk, 42.50.Lc}

\section{Introduction}	

Kenneth and Klich proved\cite{Kenneth20061} that the interaction due to vacuum fluctuations between disjoint bodies is finite. One can extend this theorem and show\cite{Schaden:2010wv} that all irreducible $N$-body contributions to the Casimir energy of a locally interacting quantum field are finite if the $N$ classical objects have no \emph{common} intersection. For more than two objects this part of the vacuum energy is finite even if they are not mutually disjoint. Ignoring the backreaction of the classical background in semi-classical approximation\cite{Weinberg:1995mt},  the total zero-point energy $\E_\set{12\dots N}$ of a quantum field that interacts locally with $N$-objects may be decomposed into irreducible contributions $\tE_s$ of the distinct subsets $s\subseteq\{12\dots N\}$ of objects,
\bel{tE}
\E_\set{12\dots N}=\sum_{s\subseteq\set{12\dots N}} \tE_s\ .
\ee
The irreducible parts of the vacuum energy are recursively obtained as,
\begin{align}
\tE_\emptyset&=\E_\emptyset\ , \nonumber\\
\tE_\set{i}&=\E_\set{i}-\E_\emptyset\ , \nonumber\\   
\tE_\set{ij}&=\E_\set{ij}-\E_\set{i}-\E_\set{j}+\E_\emptyset\ , \nonumber\\
\tE_\set{ijk}&=\E_\set{ijk}-\E_\set{ij}-\E_\set{ik}-\E_\set{jk}+\E_\set{i}+\E_\set{j}+\E_\set{j}-\E_\emptyset\ , \ \text{etc.}
\label{tEsol}
\end{align}
Although the individual vacuum energies on the rhs of~\equ{tEsol} generally diverge, the irreducible many-body contributions on the lhs of~\equ{tEsol} have a finite limit as the regularization is removed if the common intersection of the $N$ objects is empty. Note that the irreducible $N$-body Casimir energy depends crucially on what one considers to be the $N$ objects. In general it does not coincide with the work needed to assemble the $N$ objects from infinity. We here consider only a massless scalar quantum field. For proofs in the more general case of a massless bosonic field interacting locally with the objects see ref.~[\refcite{Schaden:2010wv}].

\section{Irreducible Many-Body Casimir Energies of a Massless Scalar Field}
\label{theorems}
The definition of irreducible $N$-body Casimir energies by~\equ{tE} implicitly assumes a UV-regularization of the theory. One can avoid any discussion of this regularization procedure by instead considering the intrinsically finite spectral function $\phi_{\D_s}(\beta)$ (or single-particle partition function, or trace of the heat kernel $\mathfrak{K}_{\D_s}$) of the negative Laplace-operator on the finite domain $\D_s$ containing the objects in the set $s$,
\bel{spfunc}
\phi_s(\beta)=\Tr\mathfrak{K}_{\D_s}(\beta)=\sum_{n\in\mathbb{N}} e^{-\beta\lambda_n(\D_s)/2}.
\ee

$\D_s$ is the domain $\D_\emptyset$ with objects $\set{O_j;j\in s}$ embedded; $\D_{1\dots N}$ representing the \emph{finite} domain $\D_\emptyset$ with all $N$ objects included.  $\set{\lambda_n(\D_s)>0,n\in \hbox{\text{I\hspace{-2pt}N}}}$ is the discrete spectrum of a massless scalar field that vanishes on the boundary of $\D_\emptyset$ and whose interactions with the objects in $\D_s$ are \emph{local}. We assume that the interaction of scalar field interacts with the $k^\text{th}$ object is described by a positive local potential $V_k(\bx)$ whose support is the domain $O_k$ defining the object. This precludes Neumann boundary conditions but does include the possibility of imposing Dirichlet boundary conditions on an object\footnote{The latter are enforced in the limit $V_k(\bx\in O_k)\rightarrow\infty$}.

Regularized vacuum energies $E_s^{(\eps)}$ for the domain $\D_s$ may, for instance, be defined in terms of spectral functions by introducing a proper-time cutoff $\eps$,
\bel{casE}
\E_s^{(\eps)}=-\frac{\hbar c}{\sqrt{8\pi}}\int_\eps^\infty \phi_s(\beta)\frac{d\beta}{\beta^{3/2}}\  .
\ee
The regularized vacuum energies $E_s^{(\eps)}$ generally diverge in the limit $\eps\rightarrow 0^+$ due to the asymptotic behavior of $\phi_s(\beta)$\cite{Greiner711,Gilkey841,kirsten2002spectral,Vassilevich20031},
\bel{highT}
\phi_s(\beta\sim 0)\sim \sum_{\nu=-d}^\infty (2\pi\beta)^{\nu/2} A^{(\nu)}_s+\O(e^{-\ell_\text{min}^2/(2\beta)})\ .
\ee
\Equ{highT} is given for $d$ spatial dimensions and the Hadamard-Minakshisundaram-DeWitt-Seeley coefficients $A^{(\nu)}_s$  have canonical mass-dimension $\nu$.  They reflect geometric properties\cite{Kac:1966xd} of the domain $\D_s$. The exponentially suppressed terms in the asymptotic expansion of~\equ{highT} are associated with classical periodic paths of minimal length $\ell_{min}$. The regularized vacuum energy defined by~\equ{casE} diverges only if one of the first $d+2$ coefficients of the asymptotic heat kernel expansion does not vanish. By choosing a suitable linear combination of a finite number of spectral functions to different domains, the first few offending coefficients in the asymptotic high-temperature $(\beta\sim 0)$ may be canceled\cite{Schaden20091} so that the corresponding linear combination of vacuum energies remains finite for $\varepsilon\rightarrow 0^+$. 

\begin{figure}
\includegraphics[width=\textwidth]{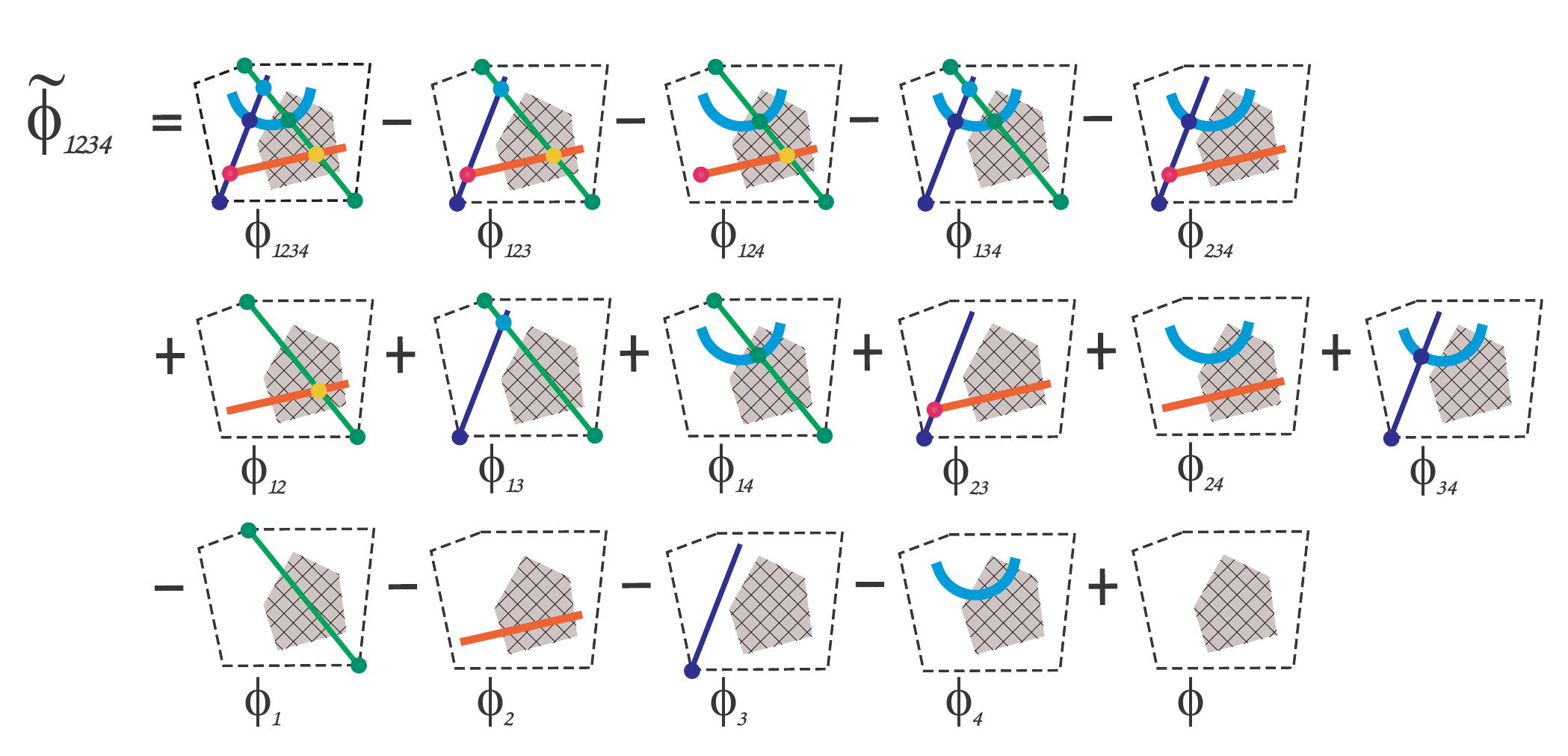}
\vspace*{8pt}
\caption{\small The subtracted spectral function $\tphi_\set{1234}(\beta)$ defined in~\equ{tphi} for a bounded two-dimensional domain $\D_\emptyset$ with four intersecting curve segments as objects. Each pictograph represents the spectral function of the domain taken with the indicated sign. Curves of different color correspond to domains of possibly different local potentials.   A random walk that crosses only three of the four segments is shown schematically. As explained in the text, it does not contribute to the Feynman-Kac integral representation of this irreducible spectral function.}
\label{fourline}
\end{figure}

Let us therefore define an irreducible spectral function for the set of objects in $s$ corresponding to the linear combinations in~\equ{tEsol},    
\bel{tphi}
\tphi_s(\beta):=(-1)^{|s|}\sum_{r\subseteq s} (-1)^{|r|}\phi_r(\beta)\ ,
\ee 
where $|r|$ denotes the cardinality of the finite set $r$ and the sum extends over all distinct subsets of objects, $r\subseteq s$.   A pictorial description of~\equ{tphi} for a set of four curve segments in a bounded two-dimensional Euclidean space $\D_\emptyset$  is given by~\fig{fourline}.

Although this holds more generally\cite{Schaden:2010wv}, the Feynman-Kac theorem\cite{Feynman481,Feynman19651,Kac:1966xd,Gies20031,Gies20061,Gies20062}  for a scalar field with potential interactions implies  that the asymptotic power expansion of $\tphi_s(\beta)$ vanishes to \emph{all} orders in $\beta$ if the \emph{common} intersection of the objects in the set $s$ is empty. The argument is as follows\cite{Schaden:2010wv}.

For a massless scalar with potential interactions the Feynman-Kac theorem gives a representation of the spectral function of~\equ{spfunc} by random walks,
\bel{support}
\phi_s(\beta)=\int_{\D_\emptyset} \frac{d\bx}{(2\pi \beta)^{d/2}} \P_{\D_s}[\bell_\beta(\bx)]\ ,
\ee
where $\P_{\D_s}[\bell_\beta(\bx)]$ is the probability for a standard Brownian bridge\footnote{A standard Brownian bridge $\bell_\beta(\bx)=\{\bx+\sqrt{\beta}(\mathbf{W}(\tau)-\tau\mathbf{W}(1)); 0\leq \tau\leq 1\}$ is generated by a  $d$-dimensional Wiener process with stationary and independent increments for which $\mathbf{W}(\tau>0)$ is normally distributed with variance $\tau d$ and vanishing average\cite{Karatzas1991bk}.}, $\bell_\beta(\bx)=\{\bx_\tau, 0\leq \tau\leq\beta; \bx_0=\bx_\beta=\bx\}$, that starts at $\bx$ and returns to $\bx$ after "time" $\beta$, to not exit $\D_s$ and to survive.

The probability $p_s(\bell_\beta(\bx))$ of a particular bridge $\bell_\beta(\bx)$ to survive in $\D_s$ is, 
\bel{survival}
p_s(\bell_\beta(\bx))=\exp[-\int_0^\beta \V_s(\bx_\tau) d\tau]\ , 
\ee
where $\V_s(\bx)=\sum_{k\in s} V_k(\bx)$ is the sum of local potentials representing the objects in $\D_s$. [Dirichlet boundary conditions are imposed by setting $p_s=0$ if the loop crosses the surface of any object and $p_s=1$ if it does not.]

The contribution to $\tphi_s(\beta)$ of a loop $\ell_\beta(\bx)$ that remains within $\D_\emptyset$ and encounters all objects of the proper subset $r\subsetneq s$ and no others is,
\bel{loopcont}
\sum_{u\subseteq s} p_{u\cap r }(-1)^{|s|-|r|}=\sum_{u\subseteq r}p_u (-1)^{|s|-|u|} \sum_{n=0}^{|s|-|r|}\hspace{-1em}(-1)^{n}\genfrac{(}{)}{0pt}{}{|s|-|r|}{n}=0\ ,
\ee
It does not contribute to $\tphi_s(\beta)$ whatever the survival probabilities $p_u$. Only loops that touch all objects in $s$, ($r=s$) contribute to the alternating sum in~\equ{tphi} and we have that,
\bel{pathtphi}
\tphi_s(\beta)=(-1)^{|s|}\int_{\D_\emptyset} \frac{d\bx}{(2\pi \beta)^{d/2}} \tilde\P_{\D_s}[\bx;\beta]\ ,
\ee
where $\tilde\P_{\D_s}[\bx;\beta]$ is the probability that a standard Brownian bridge starting at $\bx$ and returning to $\bx$ after "time" $\beta$ does not exit $\D_\emptyset$ and is killed by \emph{every} one of the  objects in $s$. 

A Brownian bridge is killed by every object in $s$ with probability,
\bel{killprob}
p(\text{killed by \emph{every} objects in $s$})=\sum_{r\subseteq s}(-1)^{|r|}p_r\ .
\ee
~\equ{killprob} is the extension to $|s|$ objects of the statement,
\begin{align}
\label{composeprob}
p(\text{killed by $O_1$ \emph{and} killed by $O_2$})&=(1-p_\set{1})+(1-p_\set{2})-(1-p_\set{12})\nonumber\\
&=p_\emptyset-p_\set{1}-p_\set{2}+p_\set{12}.
\end{align}
Note that eqs.~(\ref{pathtphi})~and~(\ref{killprob}) do not require survival probabilities to be independent ($p_\set{12}=p_\set{1}p_\set{2}$), which is true only for mutually disjoint objects described by potentials with non-overlapping support.

If the objects in $s$ have no common intersection, the shortest path connecting all the objects has a finite minimal length $\ell_\text{min}>0$ and the conditional probability $\tilde\P_{\D_s}[\bx;\beta]$ is bounded by
\bel{bound} 
\tilde\P_{\D_s}[\bx;\beta]\leq e^{-\ell^2_\text{min}/(2\beta)}\ .
\ee
We thus have the 
\begin{theorem}
\label{finiteness}
Let $\D_\emptyset$ be a flat $d$-dimensional Euclidean domain of finite volume and $\V_s(\bx)=\sum_{k\in s} V_k(\bx)$ a non-empty finite set of positive potentials in $\D_\emptyset$ with supports $\text{supp}(V_k)=:O_k\subset \D_\emptyset$. If the common support $O_1\cap O_2\cap\dots\cap O_{|s|}=\emptyset$, then the irreducible spectral function $\tphi_s(\beta)$ of a scalar field interacting with $\V_s(\bx)$  [defined in eqs.~(\ref{spfunc})~and~(\ref{tphi})] satisfies,\\
\begin{equation}
0\leq \lim_{\beta\rightarrow 0^+}\beta^{d/2)} e^{\ell^2_\text{min}/(2\beta)}\tphi_s(\beta)< K 
\end{equation}
for some finite $K\leq\infty$ and length $\ell_\text{min}>0$. $\ell_\text{min}$ is bounded below by the length of the shortest classical periodic orbit that touches all the supports $O_k, k=1,\dots,|s|$.
\end{theorem}
A consequence of Theorem~\ref{finiteness} is 
\begin{corollary}
\label{asymcoro1}
If the conditions of  Theorem~\ref{finiteness} are satisfied, the asymptotic heat-kernel expansion,
\bel{astphi}
\tphi_s(\beta\sim 0)\sim \sum_{\nu=-d}^\infty (2\pi\beta)^{\nu/2} \tilde A^{(\nu)}_s+\O(e^{-\ell_\text{min}^2/(2\beta)})\ ,
\ee 
vanishes to all orders in $\beta$, that is  $\tilde A^{(\nu)}_s=0,\forall \nu$.
\end{corollary}
The asymptotic behavior implied by Theorem~\ref{finiteness} together with the fact that the spectrum is strictly positive if Dirichlet boundary conditions are imposed on the boundary of the finite domain $\D_\emptyset$ finally leads to the statement,
\begin{corollary}
\label{asymcoro2} 
if the conditions of  Theorem~\ref{finiteness} are satisfied, the irreducible many-body Casimir energy, 
 \begin{equation}
 \label{reltEtphi}
\tE_s =-\frac{\hbar c}{\sqrt{8\pi}}\int_0^\infty \tphi_s(\beta)\frac{d\beta}{\beta^{3/2}}\  ,
\end{equation}
is finite, $|\tE_s|<\infty$.
\end{corollary}
 
Furthermore, since $\tilde\P_{\D_s}$ is a positive probability, the factor of $(-1)^{|s|}$ in~\equ{pathtphi} determines the sign of $\tphi_s(\beta)$ and of $\tE_s$.  This is most readily seen for Dirichlet boundary conditions on all objects, since the sign in this case arises because paths that touch all objects in $s$ contribute only to $\phi_\emptyset(\beta)$. For a scalar field interacting locally with potentials, we thus in addition have that,
\begin{theorem}
\label{sign}
If the conditions of Theorem~\ref{finiteness} are satisfied, the irreducible many-body spectral function of a scalar field, $\tphi_s(\beta)$ defined by~\equ{tphi},  is up to a sign the conditional probability that a standard random walk,
\begin{romanlist}[ii)]
\item returns to its starting point $\bx$ after time $\beta$ and
\item never exits $\D_\emptyset$ and
\item is killed by every object in $s$.
\end{romanlist}
The sign of $\tphi_s(\beta)$  does not depend on $\beta$ and is positive for an even and negative for an odd number of objects. \Equ{reltEtphi} then implies that,
\bel{signE}
(-1)^{|s|}\tE_s<0.
\ee
\end{theorem}

It perhaps is remarkable that the sign of $\tE_s$ in the scalar case depends on the number of objects only: for scattering by local potentials, the irreducible scalar two-body Casimir energy, in particular, is negative.\Equ{signE} also holds in the limit of Dirichlet boundary conditions. 

The fact that the interaction in the scalar case is represented by a conditional probability for random walks, allows one to also sharpen the theorem\cite{Kenneth20061,Bachas20071} that the Casimir interaction is attractive for two disjoint objects that are mirror images of each other.  
\begin{theorem}
\label{monotonic}
For a scalar field interacting locally with two potentials $V_1(\bx)$ and $V_2(\bx)$ whose support can be separated by an imaginary plane, the (positive) irreducible $2$-body spectral function $\tphi_{12}(\beta)$ for any given $\beta>0$ decreases monotonically with increasing separation of the objects from the imaginary plane and therefore each other (without changing the orientation of the two objects).
\end{theorem} 
In the scalar case the theorem follows from the representation of~\equ{pathtphi} and the fact that (for given $\beta$) the conditional probability for a random walk to be killed by both objects and return decreases with increasing separation from the separating plane. The probability for a random walk to survive its encounter with an object in~\equ{survival} depends only on the sections of the random walk \emph{within} the support of the corresponding potential.  Conditioning on the sections of the walk within the objects (and thus on the probability that it is killed),  the probability that a random walk encounters both objects and returns in the given time $\beta$ therefore decreases monotonically with the separation between them, leading to Theorem\ref{monotonic}.

Applying Theorem~\ref{monotonic} to calculate the force between the two objects using~\equ{reltEtphi}, greatly restricts  the Casimir force between disjoint objects in the scalar case:
\begin{corollary}
\label{attractive}
If two objects that can be unambiguously separated by an imaginary plane  the Casimir force between them due to a scalar field whose interaction with the objects is described by local positive potentials is always attractive, irrespective of their shape and any symmetry between them.
\end{corollary}     
This extension in the scalar case of the general theorem\cite{Kenneth20061,Bachas20071} based on reflection positivity holds for potential interactions only. \Equ{signE}, Theorem~\ref{monotonic} and Corollary~\ref{attractive} hold for Dirichlet- but, for instance,  would lead to erroneous conclusions for Neumann- boundary conditions\cite{Boyer:1968uf},  that are not described by positive potential interactions. 

\section{Analytic and Numerical Examples}
\label{applications}
A multiple scattering formulation of irreducible many-body Casimir energies was developed in ref.~[\refcite{Shajesh20111,Shajeshtalk20112}]. The theorems of sect.~\ref{theorems} were verified in all cases studied analytically. We also obtained\cite{Schaden:2010wv} an exact analytical expression for the many-body Casimir energy of a $d$-dimensional tic-tac-toe, consisting of $d$ mutually orthogonal pairs of $(d-1)$-dimensional parallel hyper-planes on which  Dirichlet boundary conditions were imposed. For ordinary tic-tac-toe in $d=2$-dimensions with two pairs of parallel lines this expression reduces to
\bel{rectangle}
\tE_\#=-\frac{\hbar c }{8\pi}\sum_{n_1=1}^\infty\sum_{n_2}^\infty\frac{{\cal A}_\#}{[(n_1 w)^2+(n_2 h)^2]^{3/2}},
\ee
where  ${\cal A}_\#= w\times h$ is the area of the rectangle enclosed by the four lines and $L(n_1,n_2)=2\sqrt{(n_1 w)^2+(n_2 h)^2}$ is the length of a classical periodic orbit in its \emph{interior} that reflects $n_1$ times off the pair of lines parallel to the $y$-axis that are a distance $w$ apart and $n_2$ times off the pair of lines parallel to the $x$-axis that are a distance $h$ apart. Only classical periodic orbits that touch all four lines contribute to $\tE_\#$.  Consistent with the arguments of sect.~\ref{theorems},  the irreducible four-body  Casimir energy $\tE_\#$ is negative and finite and remains so in the limit where one of the dimensions of the rectangle vanishes and two parallel lines coincide.  When \emph{any} dimension of the rectangle becomes large, $\tE_\#$ vanishes. This analyticity in the position of the objects is expected from the world-line description. It is one of the more interesting characteristics of the irreducible many-body Casimir energies defined by  eqs.~(\ref{reltEtphi})~and~(\ref{pathtphi}). 

\Equ{rectangle} may be derived by enclosing the tic-tac-toe in a large rectangular domain $\D_\emptyset$ and computing the spectral functions $\phi_s$ on the rhs of~\equ{tphi}. For Dirichlet boundary conditions, every one of these is given by the product of spectral functions for the individual rectangular regions forming the domain. It is straightforward to show that only the part of the contribution from the enclosed rectangle given by~\equ{rectangle} survives in the alternating sum of the irreducible $4$-body Casimir energy $\tE_\#$.  Rather than showing this more explicitly, we here will compare~\equ{rectangle} with a numerical computation of $\tE_\#$ based on~eqs.~(\ref{reltEtphi})~and~(\ref{pathtphi}). 

Substituting the~\equ{pathtphi} in~\equ{reltEtphi} gives for $N$ objects in $d=2$ dimensions,
\bel{full}
\tE_\D=-\frac{\hbar c (-1)^N}{2 (2\pi)^{3/2}}\int_0^\infty \frac{d\beta}{\beta^{5/2}}\int_{\mathcal{R}^2} d\bx \tilde\P_\D[\bx;\beta]
\ee
where we have taken the limit $\D_\emptyset\rightarrow \mathcal{R}^2$.  For Dirichlet boundary conditions on the (now infinite) lines  $\tilde\P_\D[\bx;\beta]$ is the probability of a standard Brownian bridge that returns to $\bx$ after time $\beta$ to cross all the planes in $\D$. It is numerically advantageous to use the translation and scaling property of Brownian bridges,
\bel{scale}
 \P_{\mathcal{R}^2}[\bell_\beta(\bx)]= \P_{\mathcal{R}^2}[\sqrt{\beta}\bell_1(0)]\  ,
\ee 
where $\sqrt{\beta}\bell_1(0)$ denotes a unit standard Brownian bridge over the time interval $\beta=1$ starting and ending at the origin whose points have been rescaled by $\sqrt{\beta}$. It therefore suffices to numerically generate standard unit loops $\bell_1(0)$ with the Wiener measure\cite{Karatzas1991bk}. The problem at hand furthermore is \lq\lq{}convex\rq\rq{} in the sense that any loop rescaled by a factor larger than one crosses all the lines if the loop itself does. For convex problems like this, ~\equ{full} may be written in the form,
\bel{fullnum}
\tE_\D=-\frac{\hbar c (-1)^N}{2 (2\pi)^{3/2} M}\sum_{n=1}^M w(\bell^{(n)};\D), \  \text{with}\ w(\bell;\D)=\int_{\beta_0(\bell)}^\infty \frac{d\beta}{\beta^{5/2}} {\cal A}(\sqrt{\beta}\bell;\D)\ ,
\ee
where $M$ is the total number of unit standard Brownian bridges used in the numerical evaluation and $w(\bell;\D)$ is the weight of the particular unit standard Brownian bridge $\bell$. The weight is determined  by the area of support, ${\cal A}(\sqrt{\beta}\bell;\D)$, of the spatial integral in~\equ{full} for which the rescaled unit loop $\sqrt{\beta}\ell(\bx)$ crosses all planes in $\D$, $\beta_0(\ell)$ being the minimal scale for which this occurs.

\begin{figure}
\begin{minipage}{\textwidth}
\begin{minipage}{0.49\textwidth}
\includegraphics[width=\textwidth]{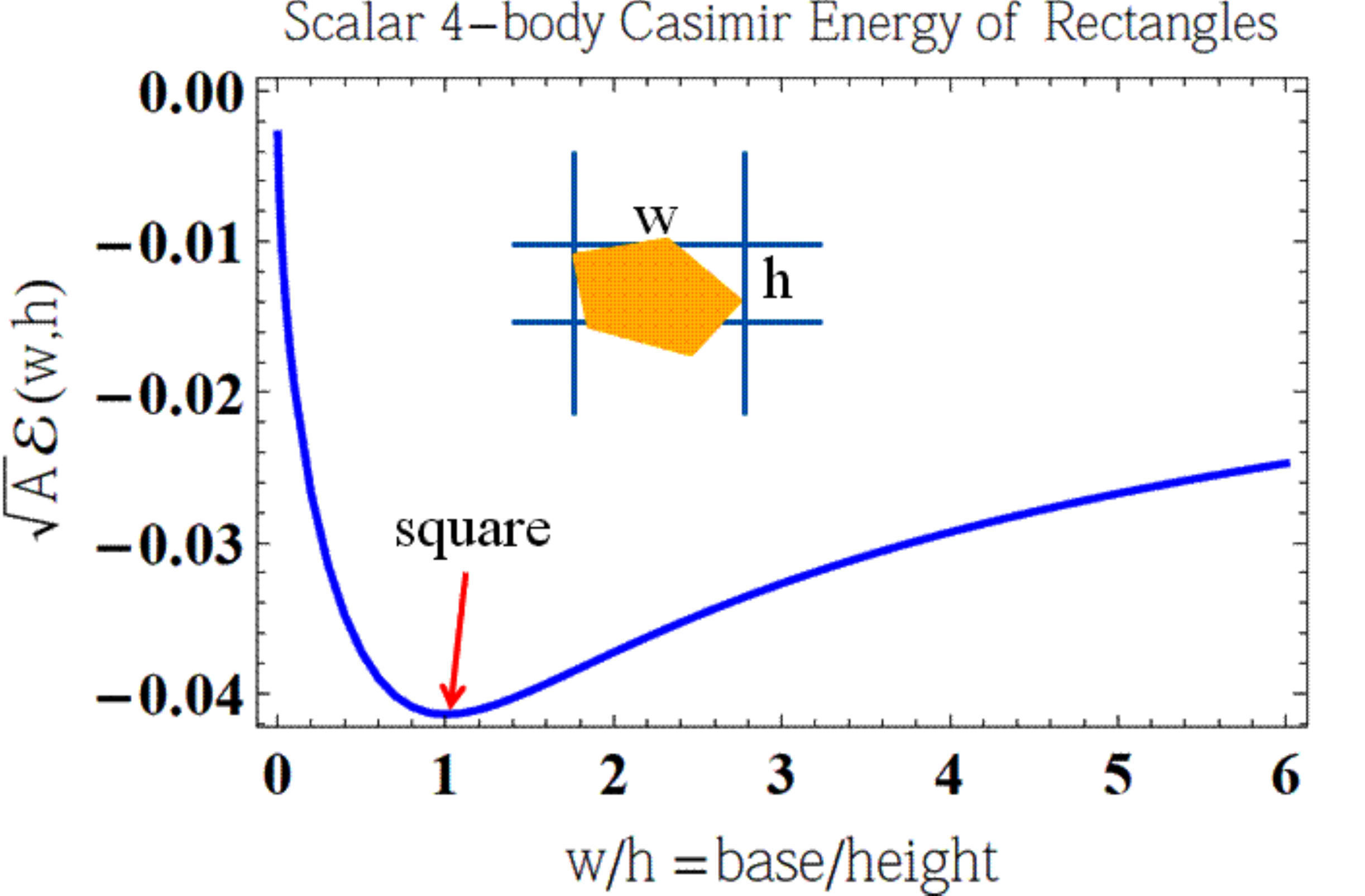}
\label{Rectanglefig}
\end{minipage}
\begin{minipage}{0.49\textwidth}
\includegraphics[width=\textwidth]{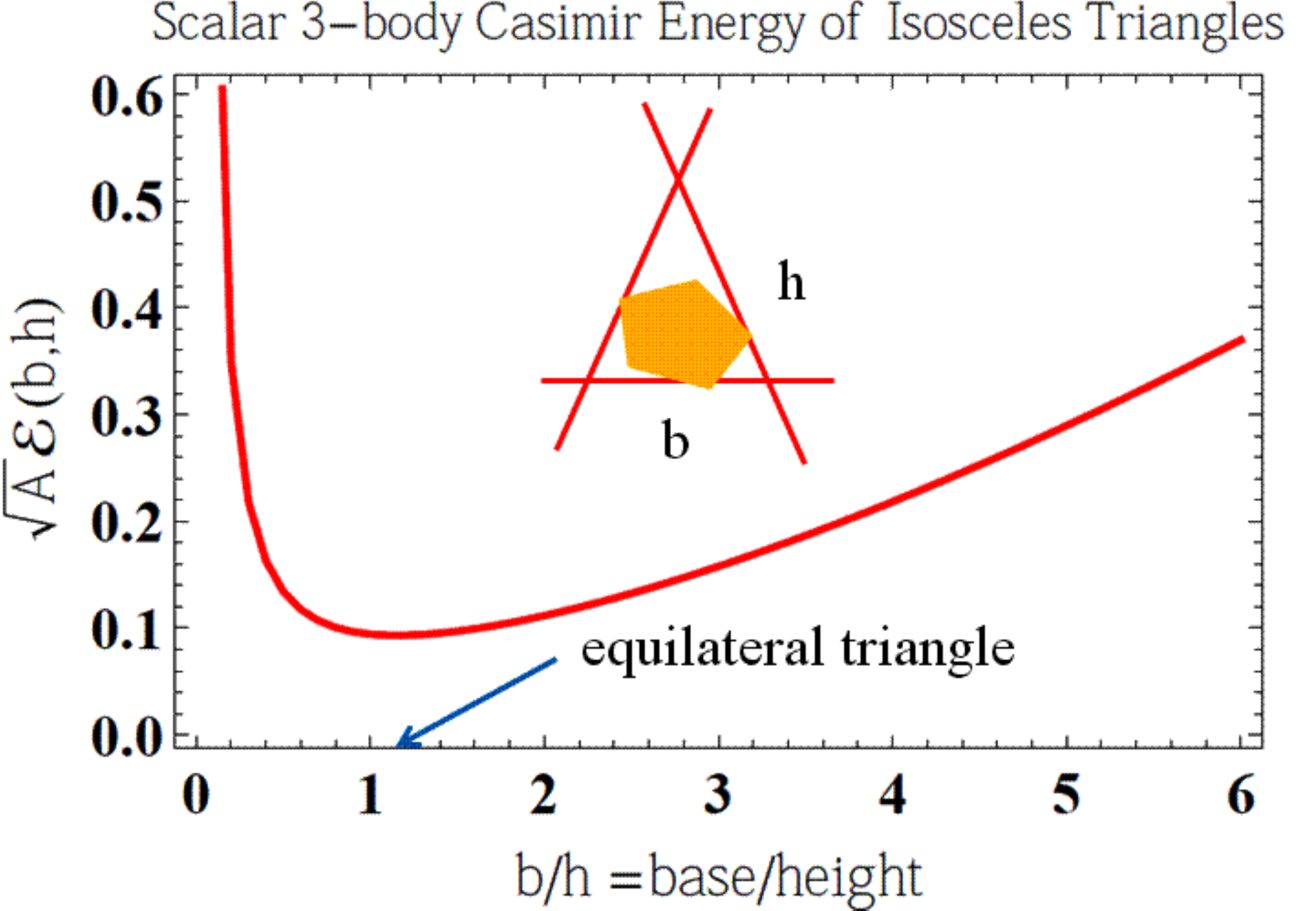}
\label{trianglefig}
\end{minipage}
\end{minipage}
\caption{{\small\linespread{0.5} Irreducible $4$- and $3$-body Casimir energies, respectively, for a massless scalar field satisfying Dirichlet boundary conditions on four lines in a tic-tac-toe pattern (left) and on three intersecting lines forming an isosceles triangle (right). The dimensionless Casimir energies scaled by the square-root of the area ${\cal A}$ of the enclosed figure are shown as functions of the base-to-height ratio of the enclosed figure. Inserts sketch the geometry of the line configurations and a schematic hull of a Brownian bridge that contributes to the irreducible Casimir energy because it intersects all objects.}}
\label{polytopes}
\end{figure}

For a tic-tac-toe configuration enclosing a rectangular region $w\times h={\cal A}_\#$ and a standard Brownian bridge $\ell_1(0)$ whose (greatest) extent in $x$- and $y$-directions is $\Delta x$ and $\Delta y$, respectively, the weight in~\equ{fullnum} is given by,
\bel{tic-tac-toe}
w(\bell({\Delta x,\Delta y}); \#_{w\times h})=\frac{s_\text{min}^2( s_\text{max}- \frac{s_\text{min}}{3})}{\sqrt{\ell_x\ell_y}}\  ;\  
s_{\genfrac{}{}{0pt}{}{\text{min}}{\text{max}}} = \genfrac{}{}{0pt}{1}{\text{min}}{\text{max}} \left\{\Delta x\sqrt{\frac{h}{w}},\Delta y\sqrt{\frac{w}{h}}\right\}
\ee
The corresponding four-body irreducible contribution to the Casimir energy,
\bel{Castic-tac-toe}
\tE_\#(w,h)=-\frac{\hbar c }{2 (2\pi)^{3/2} \sqrt{{\cal A}_\#}}E[s_\text{min}^2( s_\text{max}- \frac{s_\text{min}}{3})]=\frac{\hbar c }{\sqrt{{\cal A}_\#}} \varepsilon_\#(w/h)\ ,
\ee
is proportional to the expectation $E[s_\text{min}^2( s_\text{max}- \frac{s_\text{min}}{3})]$ for extremal values of standard Brownian bridges under the Wiener measure.  We used a set of 1,000 unit standard Brownian bridges\footnote{To improve numerical accuracy we make use of the rotational invariance of the Wiener measure and include up to 6 rotated duplicates for each generated Brownian bridge.} to calculate $\varepsilon_\#$.  The result is shown in   Fig.~\ref{polytopes}a and coincides with that obtained by numerical evaluation of the sum in~\equ{rectangle} to within 1\%.
 
For a set of three lines whose intersection forms a triangle with area ${\cal A}_\triangle$, the weight is entirely determined by the minimal scale $\beta_0(\ell)$ for which the standard Brownian bridge just touches all three lines,
\bel{triangle}
w(\bell; \triangle )= \frac{2{\cal A}_\triangle}{3 \beta_0^{3/2}(\ell)}.
\ee
Note that the minimal scale $\beta_0(\ell)$ in~\equ{triangle} has the dimension (length)$^2$. Inserting~\equ{triangle} in~\equ{fullnum}, the irreducible three-body Casimir energy for a configuration of lines forming an isosceles triangle of height $h$ and base $b$ with area ${\cal A}_\triangle= h b/2$ is,
\bel{Castriangle}
\tE_{\text{iso}\triangle}(h,b)=\frac{\hbar c {\cal A}_\triangle}{3 (2\pi)^{3/2}}E[{\beta_0^{-3/2}(\ell)]}=\frac{\hbar c}{\sqrt{{\cal A}_\triangle} }\varepsilon_{\text{iso}\triangle}(b/h),
\ee
where the expectation $E[{\beta_0^{-3/2}(\ell)]}$ again is with respect to the Wiener measure for unit standard Brownian bridges. Using the same set of Brownian bridges as for the tic-tac-toe, the numerical result is shown in Fig.~\ref{polytopes}b. Note that,  as for the tic-tac-toe, the irreducible $3$-body Casimir energy also is minimal at the symmetric configuration. However, whereas the irreducible Casimir energy of the tic-tac-toe is negative and finite when two of the four lines coincide, the irreducible Casimir energy of the triangular configuration is strictly positive and diverges when its three sides coincide.

\section*{Acknowledgments}
I would like to thank the organizers of QFEXT11 for a very interesting and engaging conference.  This work was supported by the National Science Foundation with  Grant no.~PHY0902054.
\bibliographystyle{unsrt}

\bibliography{c:/Users/Martin/Rutgers/research/casimir/bibliothek/b20111010-proposal,c:/Users/Martin/Rutgers/research/casimir/bibliothek/b20111010-proposal-extra,c:/Users/Martin/Rutgers/research/casimir/bibliothek/b7001-many-body,c:/Users/Martin/Rutgers/research/casimir/bibliothek/b20100712-roughness,c:/Users/Martin/Rutgers/research/casimir/bibliothek/b7002-Martins-data,c:/Users/Martin/Rutgers/research/casimir/bibliothek/b01-schwinger,c:/Users/Martin/Rutgers/research/casimir/bibliothek/b05-plates,c:/Users/Martin/Rutgers/research/casimir/bibliothek/b02-fall-2,c:/Users/Martin/Rutgers/research/casimir/bibliothek/b8001-thin-plate,c:/Users/Martin/Rutgers/research/casimir/bibliothek/b04-gears-1}
\end{document}